\documentclass[aps,showpacs,prl,twocolumn,groupedaddress]{revtex4-1}

\usepackage{graphicx}
\usepackage{natbib}

\begin{document}


\title{Magnetic vortex echoes: application to the study of arrays of magnetic nanostructures}



\author{F. Garcia$^{1}$}
\author{J.P. Sinnecker$^{2}$}
\author{E.R.P. Novais$^{2}$}
\author{A.P. Guimar\~aes$^{2}$}

\email[Author to whom correspondence should be addressed:]{ apguima@cbpf.br}

\affiliation{$^{1}$Laborat\'orio Nacional de Luz S\'{\i}ncrotron, 13083-970, Campinas, SP, Brazil}

\affiliation{$^{2}$Centro Brasileiro de Pesquisas F\'{\i}sicas, 22290-180,  Rio de Janeiro, RJ, Brazil }%



\date{\today}

\begin{abstract}
We propose the use of the gyrotropic motion of vortex cores in nanomagnets to produce a magnetic echo, analogous to the spin echo in NMR. This echo occurs when an array of nanomagnets, e.g., nanodisks, is magnetized with an in-plane ($xy$) field, and after a time $\tau$ a field pulse inverts the core magnetization; the echo is a peak in $M_{xy}$ at $t=2\tau$. Its relaxation times depend on the inhomogeneity, on the interaction between the nanodots and on the Gilbert damping constant $\alpha$. Its feasibility is demonstrated using micromagnetic simulation. To illustrate an application of the echoes, we have determined the inhomogeneity and measured the magnetic interaction in an array of nanodisks separated by a distance $d$, finding  a $d^{-n}$ dependence, with $n\approx4$.
\end{abstract}



\maketitle


The interest in magnetic vortices and their properties and applications has grown steadily in the last years\cite{Guimaraes:2009,Chien:2007,Guslienko:2008,Garcia:2010}. Vortices have been observed, for example, in disks and ellipses having  sub-micron dimensions\cite{Novais:2011}. More recently, the question of the intensity of the coupling between neighbor disks with magnetic vortex structures has attracted an increasing interest\cite{Vogel:2010,Jung:2011,Sugimoto:2011,Sukhostavets:2011}.

At the vortex core the magnetization points perpendicularly to the plane: this characterizes its polarity, $p=+1$ for the $+z$ direction and $p=-1$ for $-z$.
The direction of the moments in the vortex defines the circulation: $c=-1$ for clockwise (CW) direction, and $+1$ for CCW. If removed from the equilibrium position at the center of the nanodisk by, for example, an in-plane field, and then left to relax, a vortex core will perform a gyrotropic motion, with angular frequency $\omega$, given for thin disks\cite{Guslienko:2008} by $\omega_G\approx(20/9)\gamma M_s \beta$ ($\beta=h/R$ is the aspect ratio)\footnote{The sources of inhomogeneity are the spread in radii, in thickness or the presence of defects. An external perpendicular field  $H$ adds a contribution to $\omega$\cite{Loubens:2009}, $\omega=\omega_G + \omega_H$, with $\omega_H=\omega_0 \ p (H/H_s)$, where $p$ is the polarity and $H_s$ the field that saturates the nanodisk magnetization. A distribution  $\Delta H$ is another source of the spread $\Delta\omega$\label{NoteOmega}}.

We propose in this paper that, manipulating the dynamic properties of the vortex in an analogous way as it is done in Nuclear Magnetic Resonance (NMR), a new phenomenon results, the magnetic vortex echo (MVE), similar to the spin echo observed in NMR\cite{Hahn:1950}. This new echo may provide information on fundamental properties of arrays of nanodisks, e.g., their inhomogeneity and interactions. Despite the fact that applications of vortices necessarily involve arrays, most of the recent publications deal with the analysis of individual nanodisks or arrays with a few elements. Therefore, the possibility of characterizing large arrays is of much interest.
 In this paper we have examined the motion of vortex cores in an array of nanometric disks under the influence of a pulsed magnetic field, using micromagnetic simulation.
\begin{figure}[!]
\includegraphics[width=\columnwidth]{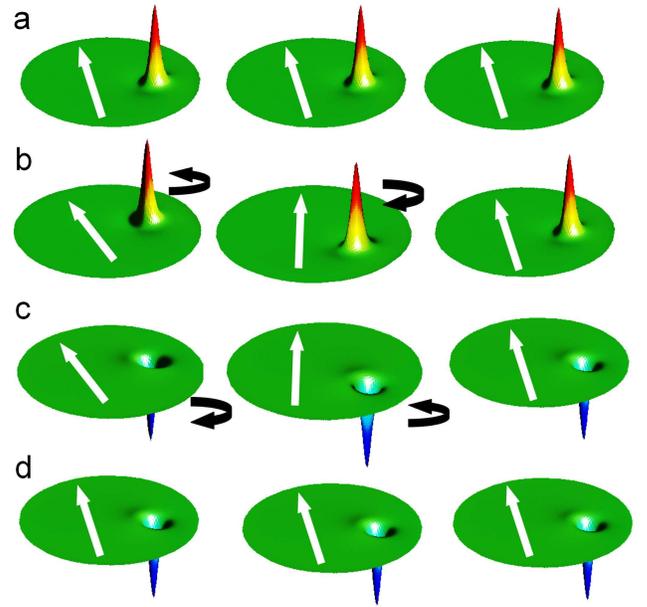}
\caption{(Color online) Diagram showing the formation of magnetic vortex echoes; the disks are described from a reference frame that turns with the average translation frequency $\omega_0$: a) disks with in-plane magnetizations ${\bf M}_i$ along the same direction (defined by the white arrows); b) after a time $\tau$ the disks on the left, center and right have turned with frequencies, respectively, lower, higher and equal to $\omega_0$; c) the polarities of the vortex cores are reversed, and the $\omega_i$ of the vortex cores (and of the ${\bf M}_i$) change sign, and d) after a second interval $\tau$ the cores (and ${\bf M}_i$) are again aligned, creating the echo.}\label{fig:EchoDiagram}
\end{figure}

Let us consider an array of nanodisks where the vortex cores precess with a distribution of angular frequencies centered on $\omega_0$, of width $\Delta \omega$, arising from any type of inhomogeneity (see note [10]). We assume that the frequencies vary continuously, and have a Gaussian distribution $P(\omega)$ with mean square deviation $\Delta\omega$.
To simplify we can assume that the polarization of every vortex is the same: $p_i=+1$. This is not necessary for our argument, but, if required, the system can be prepared; see ref. \cite{Antos:2010} and the references therein.

Since the direction of rotation of the magnetic vortex cores after removal of the in-plane field is defined exclusively by $p$, all the cores will turn in the same direction; as the vortex core turns, the in-plane magnetization of the nanodot also turns.
\begin{figure}[!]
\includegraphics[width=\columnwidth]{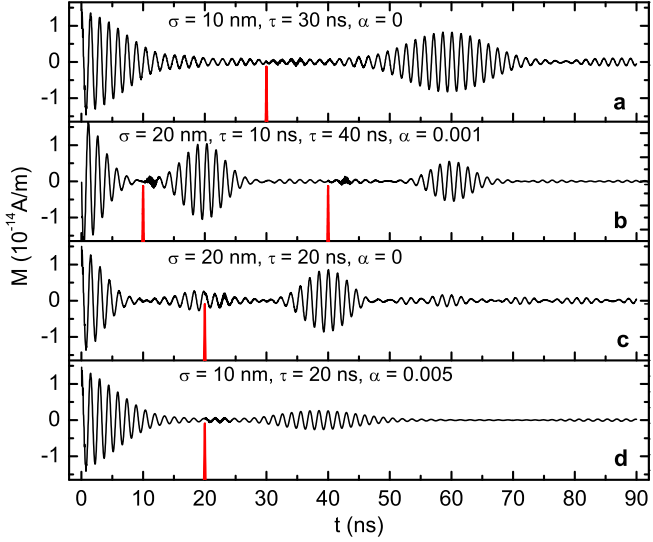}
\caption{(Color online) Micromagnetic simulation of magnetic vortex echoes, for 100 nanodisks, with $d=$ infinity, and a) $\sigma=10\,$nm, $\tau=30\,$ns, $\alpha=0$; b) $\sigma=20\,$nm, $\tau=10\,$ns and $\tau=40\,$ns (two pulses), and $\alpha=0.001$; c) $\sigma=20\,$nm, $\tau=20\,$ns, $\alpha=0$; d) $\sigma=10\,$nm, $\tau=20\,$ns, $\alpha=0.005$. The inversion pulses ($B_z=-300\,$mT) are also shown (in red).}\label{fig:EchoSamples}
\end{figure}
\begin{figure}[!]
\includegraphics[width=\columnwidth]{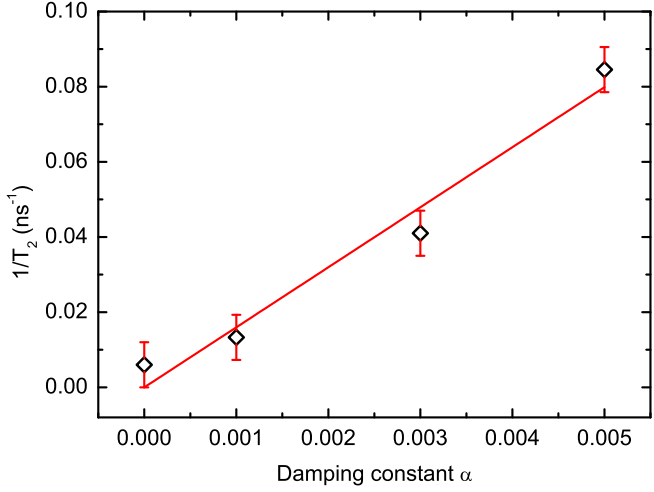}
\caption{(Color online) Variation of the inverse of the relaxation times $T_2$ (diamonds) obtained by fitting the curves of echo intensity versus time interval $\tau$ to $M_0exp\left({-{\tau}/{T_2}}\right)$, as a function of $\alpha$, for simulations made with $D=250\,$nm, $\sigma=10\,$nm, separation infinite; the continuous line is a linear least squares fit.} \label{fig:InverseT2VsAlpha}
\end{figure}

Let us assume that all the vortex cores have been displaced from their equilibrium positions along the positive $x$ axis\footnote{If the disks have different circulations ($c=\pm1$) the cores will be displaced in opposite directions, but the effect will be the same, since the ${\bf M}_i$ will all point along the same direction.}, by a field ${\bf B}$. The total in-plane magnetization (that is perpendicular to the displacement of the core) will point along the $y$ axis, therefore forming an angle $\theta_0=0$ at $t=0$. Using the approach employed in the description of magnetic resonance  (e.g., see \cite{Slichter:1990, Guimaraes:1998}), one can derive the total in-plane magnetization:
\begin{equation}
M_y(t)=M_y(0)\int_{-\infty}^{\infty}\frac{e^{\left[{-\frac{1}{2}\frac{(\omega-\omega_0)^2}{\Delta\omega^2}}\right]}}{\Delta\omega\sqrt{2 \pi}} \cos(\omega t) d\omega
\end{equation}
an integral that is\cite{Butz:2006} the Fourier transform of the function $P(\omega)$; using $T_2^{*}=1/\Delta\omega$:
\begin{equation}
M_y(t)=M_y(0) \ exp\left({-\frac{1}{2}\frac{t^2}{T_2^{*2}}}\right)\cos (\omega_0 t) \label{eq:FID}
\end{equation}
This result shows that the total magnetization tends to zero, as the different contributions to $M_y(t)$ get gradually out of phase.
This decay is analogous to the free induction decay (FID) in NMR; its characteristic time is $T_2^{*}=1/\Delta\omega$.

After a time $\tau$, the angle rotated by each vortex core will be $\omega \tau$; if at  $t=\tau$ we invert the polarities of the vortices in the array, using an appropriate pulse, the motion of the cores will change direction (i.e., $\omega \rightarrow -\omega$), and one obtains:
\begin{equation}
M_y(t-\tau)=M_y(0)\int_{-\infty}^{\infty}\frac{e^{-\frac{1}{2}\frac{(\omega-\omega_0)^2}{\Delta\omega^2}}}{\Delta\omega\sqrt{2 \pi}} \cos[\omega(\tau- t)] d\omega
\end{equation}
The magnetization at a time $t>\tau$ is then:
\begin{equation}
M_y(t)=M_y(0) \ e^{[{-\frac{1}{2}\frac{(t-2\tau)^2}{T_2^{*2}}}]}e^{({-\frac{t-\tau}{{T_2}}})}\cos(\omega_0 t)
\label{eq:FIDEcho}
\end{equation}
This means that the magnetization component $M_y(t)$ increases for $\tau<t<2\tau$, reaching a maximum at a time $t=2\tau$: this maximum is the {\it magnetic vortex echo}, analogous to the spin echo observed in magnetic resonance, which has important applications in NMR, including in Magnetic Resonance Imaging (MRI)\cite{Slichter:1990,Guimaraes:1998} (Fig. \ref{fig:EchoDiagram}).
In the case of the NMR spin echo, the maximum arises from the refocusing of the in-plane components of the nuclear magnetization.


In Eq. \ref{eq:FIDEcho} we have included a relaxation term containing the time constant $T_2$ - also occurring in NMR -, to account for a possible decay of the echo amplitude with time; its justification will be given below.

In the array of nanodisks, there will be in principle two contributions to the defocusing of the magnetization, i.e., two mechanisms for the loss of in-plane magnetization memory: 1) the spread in values of $\beta$ and $H$ (see note [10]), producing an angular frequency broadening term $\Delta \omega$, and 2) irreversible processes that are characterized by a relaxation time $T_2$: thus
${1}/T_2^{*}=\Delta \omega +  {1}/{T_2}$.

The second contribution is the homogeneous term whose inverse, $T_2$, is the magnetic vortex transverse relaxation time, analogous to the spin transverse relaxation time (or spin-spin relaxation time) $T_2$ in magnetic resonance.
The irreversible processes include a) the interaction between the disks, which amounts to random magnetic fields that will increase or decrease $\omega$ of a given disk, producing a frequency spread of width $\Delta\omega^{\prime}={1}/{T_2^{ \prime}}$, and b) the loss in magnetization  (of rate $1/T_{\alpha})$ arising from the energy dissipation related to the Gilbert damping constant $\alpha$. Identifying $T_{\alpha}$ to the NMR longitudinal relaxation time $T_1$, one has \cite{Slichter:1990}:
${1}/{T_2}= {1}/{T_2^{ \prime}} + {1}/{2T_{\alpha}}$.

Therefore the relaxation rate $1/T_2^{*}$ is given by:
\begin{equation}
\frac{1}{T_2^{*}}=\Delta \omega +  \frac{1}{T_2}=
\Delta \omega + \frac{1}{T_2^{ \prime}} + \frac{1}{2T_{\alpha}} \label{eq:T2*}
\end{equation}
$1/T_2^{*}$ is therefore the total relaxation rate of the in-plane magnetization, composed of a) $\Delta\omega$, the inhomogeneity term, and b) $1/T_2$, the sum of all the other contributions, containing $1/T_2^{\prime}$, due to the interaction between the disks, and $1/T_{\alpha}$, the rate of energy decay.
The vortex cores will reach the equilibrium position at $r=0$ after a time $t\sim T_{\alpha}$, therefore there will be no echo for $2\tau \gg T_{\alpha}$.

The vortex echo maximum at $t=2\tau$, from Eq. \ref{eq:FIDEcho}, is
$M_y(2\tau)\propto exp\left({-{\tau}/{T_2}}\right)$;
one should therefore note that the maximum magnetization recovered at a time $2\tau$ decreases exponentially with
$T_2$, i.e., this maximum is only affected by the homogeneous part of the total decay rate given by Eq. \ref{eq:T2*}. In other words, the vortex echo cancels the loss in $M_y$ due to the inhomogeneity $\Delta\omega$, but it does not cancel the decrease in $M_y$ due to the interaction between the nanodisks (the homogeneous relaxation term ${1}/{T_2^{ \prime}}$), or due to the energy dissipation (term ${1}/{2T_{\alpha}}$).

Note also that if one attempted to estimate the inhomogeneity of an array of nanodots using another method, for example, measuring the linewidth of a FMR spectrum, one would have the contribution of this inhomogeneity together with the other terms that appear in Eq. \ref{eq:T2*}, arising from interaction between the dots and from the damping. On the other hand, measuring the vortex echo it would be possible to separate the intrinsic inhomogeneity from these contributions, since $T_2$ can be measured separately, independently of the term $\Delta \omega$.
$T_2$ can be measured by determining the decay of the echo amplitude for different values of the interval $\tau$.

The Fourier transform of either the vortex free induction decay or the time dependence of the echo $M_y(t)$ gives the distribution of gyrotropic frequencies $P(\omega)$.

For the experimental study of vortex echoes, the sequence of preparation (at t=0) and inversion fields (at t=$\tau$) should of course be repeated periodically, with a period $T \gg T_{\alpha}$. As in pulse NMR, this will produce echoes on every cycle, improving the S/N ratio of the measured signals. Also note that the time $T_2^{*}$ can be obtained either from the initial decay (FID, Eq. \ref{eq:FID}) or from the echo (Eq. \ref{eq:FIDEcho}), but $T_2$ can only be obtained from the MVE.

   In order to demonstrate the MVE, we have performed micromagnetic simulations of an assembly of 100 magnetic nanodisks employing the OOMMF code\footnote{Available from  http://math.nist.gov/oommf/}. The simulated system was a square array of $10 \times 10$ disks, thickness $20\,$nm, with distance $d$ from center to center. In order to simulate the inhomogeneity of the system, we have introduced a Gaussian distribution of diameters, centered on $250\,$nm and mean square deviation $\sigma$; $\sigma=10$\,nm corresponds to  $\Delta\omega\approx 1.6 \times 10^8\,$s$^{-1}$. The disks were placed at random on the square lattice.
   The initial state of the disks ($p=+1$ and $c=-1$) was prepared by applying an in-plane field of 25$\,$mT; the polarity was inverted with a Gaussian pulse of amplitude $B_z=-300\,$mT, with width 100$\,$ps.
 The results for the case $d=\infty$ were simulations made on the disks one at a time, adding the individual magnetic moments $\mu_i(t)$.

 We have successfully demonstrated the occurrence of the magnetic vortex phenomenon, and have shown its potential as a characterization technique. The simulations have confirmed the occurrence of the echoes at the expected times ($t=2\tau$). For different values of $\sigma$, the $T_2^{*}$ time, and consequently the duration of the FID and the width of the echo are modified (Fig. \ref{fig:EchoSamples}a, \ref{fig:EchoSamples}c); increasing $\alpha$ results in a faster decay of the echo intensity as a function of time (Fig. \ref{fig:EchoSamples}a,   \ref{fig:EchoSamples}d).
We have also obtained multiple echoes, by exciting the system with two pulses (Fig. \ref{fig:EchoSamples}b)\footnote{These echoes, however, are not equivalent to the stimulated echoes observed in NMR with two 90$^o$ pulses\cite{Hahn:1950}}.

Fig. \ref{fig:InverseT2VsAlpha} shows the dependence of $T_2$ on $\alpha$ for $\sigma=10\,$nm; essentially the same result is obtained for $\sigma=20\,$nm, since $T_2$ does not depend on $\Delta\omega$ (Eq. \ref{eq:T2*}).
Taking a linear approximation, $1/T_2=A\alpha$, and since for $d=$infinity there is no interaction between the disks,  $1/T_2=1/2T_{\alpha}$, and therefore:
\begin{equation}
\frac{1}{T_{\alpha}}=2A \alpha \label{eq:Talpha}
\end{equation}
From the least squares fit (Fig. \ref{fig:InverseT2VsAlpha}), $A=1.6 \times 10^{10}\,$s$^{-1}$.
This relation can be used to determine experimentally $\alpha$, measuring $T_2$ with vortex echoes, for an array of well-separated disks.
\begin{figure}[!]
\includegraphics[width=\columnwidth]{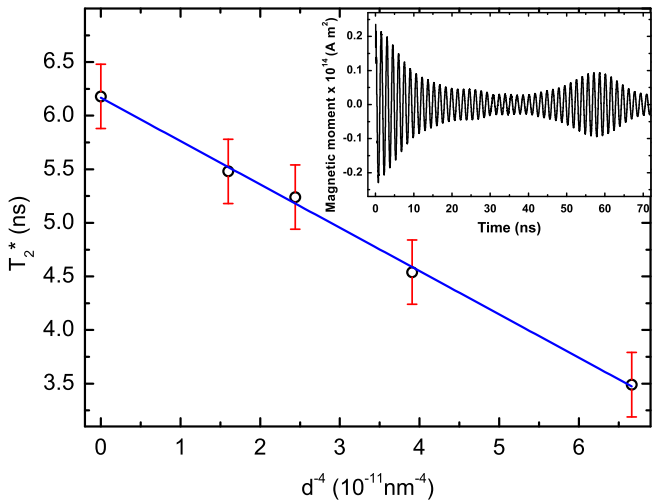}
\caption{(Color online) Variation of the relaxation times $T_2^{*}$ versus $d^{-4}$ for an array of $10 \times 10$ magnetic nanodisks with a distribution of diameters centered on $D=250\,$nm ($\sigma=10\,$nm), damping constant $\alpha=0.001$ and separation $d$ between their centers; the continuous line is a linear least squares fit. The inset shows a vortex echo simulation for the array, with $d=500\,$nm, $\tau=30\,$ns, $\alpha=0.001$.} \label{fig:RelaxVsd}
\end{figure}

 Recently some workers have analyzed the  important problem of the interaction between disks exhibiting magnetic vortices, obtaining that it varies with a  $d^{-n}$ dependence: Vogel and co-workers \cite{Vogel:2010}, using FMR, obtained for a $4\times 300$ array a dependence of the form $d^{-6}$, the same found by Sugimoto et al. \cite{Sugimoto:2011} using a pair of disks excited with rf current. Jung et al. \cite{Jung:2011} studying a pair of nanodisks with time-resolved X-ray spectroscopy, found $n=3.91\pm 0.07$
and Sukhostavets et al. \cite{Sukhostavets:2011}, also for a pair of disks, in this case studied by micromagnetic simulation, obtained $n=3.2$ and $3.7$ for the $x$ and $y$ interaction terms, respectively.

As a first approximation one can derive the dependence of the contribution to $1/T_2^{*}$ related to the distance between the disks as:

\begin{equation}
 T_2^{*}=B+C d^{-n} \label{eq:T2fit}
 \end{equation}
 From our simulations, and using Eq. \ref{eq:T2fit} we found, from the best fit, that this interaction varies as $d^{-n}$, with $n=3.9 \pm 0.1$, in a good agreement with \cite{Jung:2011} and reasonable agreement with Sukhostavets et al.\cite{Sukhostavets:2011}.

   Determining ${1}/{T_2^{ \prime}}$ has allowed us to obtain the intensity of the interaction between the disks as a function of separation $d$ between them. Substituting Eq. \ref{eq:Talpha} and Eq. \ref{eq:T2fit} in Eq. \ref{eq:T2*}, we can obtain the expression for the interaction as a function of $d$:
\begin{equation}
\frac{1}{T_2^{\prime}}=\frac{1}{B+C d^{-n}}-A\alpha-\Delta\omega
\approx\frac{d^n}{|C|}-A\alpha-\Delta\omega;
\end{equation}
(approximation valid for $d$ small). In Fig. \ref{fig:RelaxVsd} we show the results of the simulations with $\sigma=10\,$nm and $\alpha=0.001$. Assuming $n=4$ and making a linear squares fit, we obtained $B=6.15\times 10^{-9}\,$s, $C=-4.03\times 10^{-35}\,$s  m$^4$.

A new phenomenon, the magnetic vortex echo, analogous to the NMR spin echo, is proposed and demonstrated here through micromagnetic simulation. Applications of the magnetic vortex echo includes the measurement of the inhomogeneity, such as, distribution of dimensions, aspect ratios, defects, and perpendicular magnetic fields and so on, in a planar array of nanodisks or ellipses; it may be used to study arrays of nanowires or nanopillars containing thin layers of magnetic material. These properties cannot be obtained directly, for example, from the linewidth of FMR absorption.

The MVE is a tool that can be used to evaluate the interaction between the elements of a large array of nanomagnets with vortex ground states.
It can also be used to determine the Gilbert damping constant $\alpha$ in these systems.

 The authors would like to thank G.M.B. Fior for the collaboration; we are also indebted to the Brazilian agencies CNPq, CAPES, FAPERJ, FAPESP.

\bibliography{Nanomagnetism}

\end{document}